\documentclass[12pt]{article}
\usepackage[top=1.0in, bottom=1.0in, left=1.0in, right=1.0in]{geometry}
\usepackage{enumerate,fancyhdr,xcolor}
\usepackage{graphicx}
\usepackage{comment}
\begin{document} 
\setcounter{page}{1}

\begin{center}
\large{\bf White paper:\\
 CeLAND - Investigation of the reactor antineutrino anomaly with an intense $^{144}$Ce - $^{144}$Pr antineutrino source in KamLAND}\\

\vskip 0.3in
\normalsize
A. Gando, Y. Gando, S. Hayashida, H. Ikeda, K. Inoue, K. Ishidoshiro, H. Ishikawa, M. Koga, R. Matsuda, S. Matsuda, T. Mitsui,D. Motoki, K. Nakamura, Y. Oki, M. Otani, I. Shimizu, J. Shirai, F. Suekane, A. Suzuki, Y. Takemoto, K. Tamae, K. Ueshima, H. Watanabe, B.D. Xu, S. Yamada,Y. Yamauchi, H. Yoshida\\
\textit{Research Center for Neutrino Science, Tohoku University, Sendai 980-8578, Japan}

\vskip 0.1in
B. E. Berger\\
\textit{Colorado State University, Fort Collins, CO 80523-1875, USA}

\vskip 0.1in
M. Cribier${1,2}$, M. Durero$^1$, V. Fischer$^1$, J. Gaffiot$^2$,
T. Lasserre$^{1,2}$, A. Letourneau$^1$, D. Lhuillier$^1$, G. Mention$^1$, L. Scola$^1$, Ch. Veyssière$^1$ and M. Vivier$^1$\\ 
\textit{$^1$Commissariat à l’énergie atomique et aux énergies alternatives, Centre de Saclay, IRFU, 91191 Gif-sur-Yvette, France}\\
\textit{$^2$Astroparticules et Cosmologie APC, 10 rue Alice Domon et Léonie Duquet, 75205 Paris cedex 13, France}

\vskip 0.1in 
A. Kozlov\\
\textit{Kavli Institute for the Physics and Mathematics of the Universe (WPI), 
University of Tokyo, Kashiwa 277-8583, Japan}

\vskip 0.1in
T. Banks$^2$, D. Dwyer$^1$, B. K. Fujikawa$^1$, Yu. G. Kolomensky$^{1,2}$, T. O’Donnell$^2$ 
\textit{$^1$Lawrence Berkeley National Laboratory, Berkeley, CA 94720, USA}\\
\textit{$^2$University of California, Berkeley, CA 94704, USA}

\vskip 0.1in
Patrick Decowski \\
\textit{Nikhef and the University of Amsterdam, Science Park 105 1098 XG, Amsterdam, the Netherlands}

\vskip 0.1in
 D. M. Markoff\\
\textit{North Carolina Central University, Durham, NC 27707, USA}

\vskip 0.1in
S. Yoshida\\
\textit{Graduate School of Science, Osaka University, Toyonaka, Osaka 560-0043, 
Japan}

\vskip 0.1in
V.N. Kornoukhov$^1$, T. V.M. Gelis$^3$, G.V. Tikhomirov$^2$, I.S. Saldikov$^2$
\textit{$^1$ Russian Federation State Scientific Center of Theoretical and
Experimental Physics Institute, 117218 Moscow, Russia}\\
\textit{$^2$Scientific and Research Nuclear University, Moscow engineering and Physics Institute, Russia}\\
\textit{$^3$ Russian Academy of Sciences A.N. Frumkin Institute of Physical chemistry and
Electrochemistry, Russia}

\vskip 0.1in
 J. G. Learned, J. Maricic, S. Matsuno, R. Milincic\\
\textit{University of Hawaii at Manoa, Honolulu, HI 96822, USA}

\newpage
\vskip 0.1in
H. J. Karwowski\\
\textit{University of North Carolina, Chapel Hill, NC 27599, USA}

\vskip 0.1in
Y. Efremenko\\
\textit{University of Tennessee, Knoxville, TN 37996, USA}

\vskip 0.1in
J. A. Detwiler, S. Enomoto\\
\textit{University of Washington, Seattle, WA 98195, USA}
\end{center}

\abstract{We propose to test for short baseline neutrino oscillations, implied by the recent reevaluation of the reactor antineutrino flux and by anomalous results from the gallium solar neutrino detectors. The test will consist of producing a 75 kCi $^{144}$Ce - $^{144}$Pr antineutrino source to be deployed in the Kamioka Liquid Scintillator Anti-Neutrino Detector (KamLAND). KamLAND’s 13 m diameter target volume provides a suitable environment to measure energy and position dependence of the detected neutrino flux. A characteristic oscillation pattern would be visible for a baseline of about 10 m or less, providing a very clean signal of neutrino disappearance into a yet-unknown, “sterile” state. Such a measurement will be free of any reactor-related uncertainties. After 1.5 years of data taking the Reactor Antineutrino Anomaly parameter space will be tested at $>$ 95\% C.L.}

\section{Introduction} 
\begin{figure}[htb][]
\centering
\includegraphics[scale=0.2]{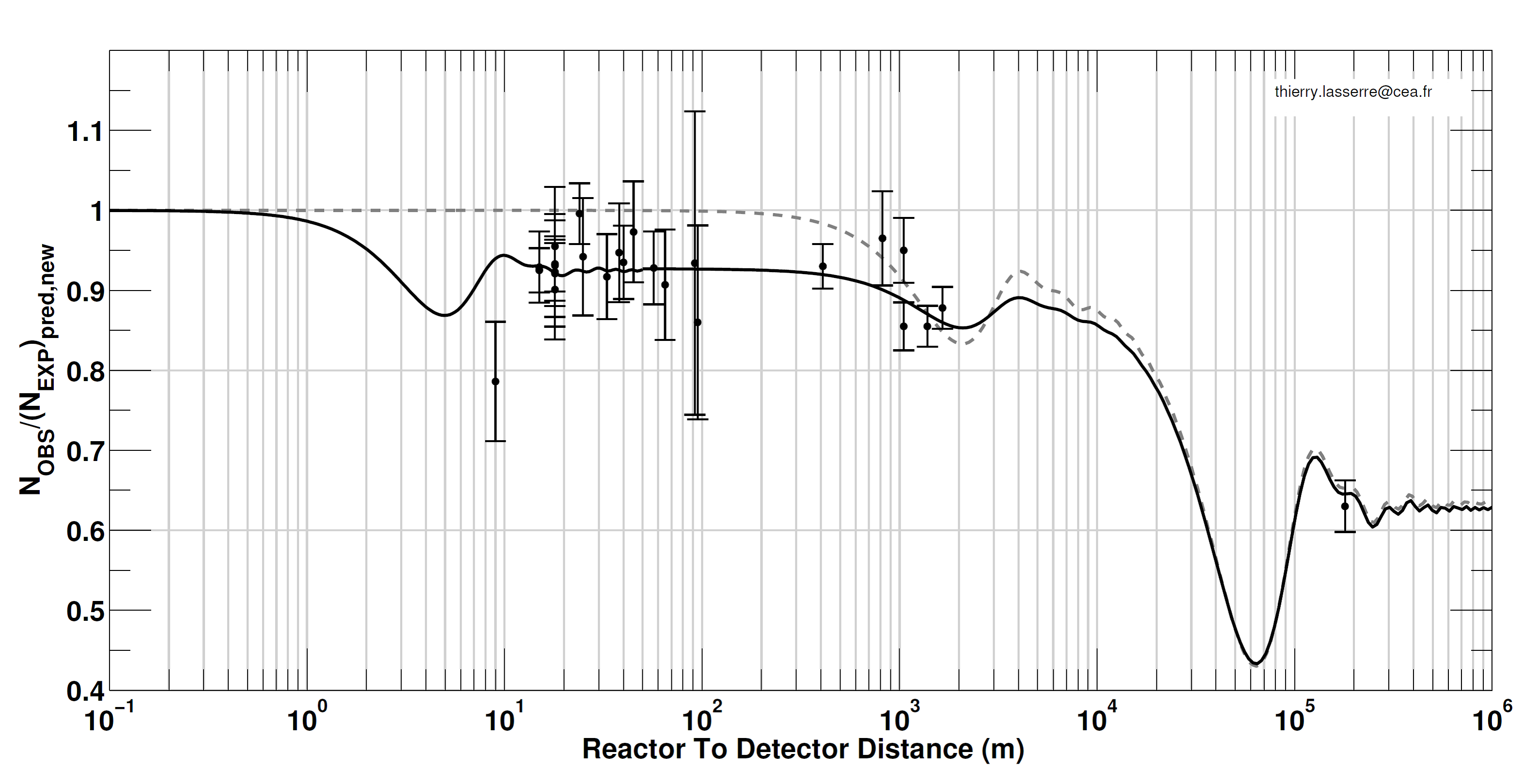}
\caption{Graph showing the ratio of the observed to the expected rate, from the recent reactor neutrino flux calculations without neutrino oscillations, for all reactor neutrino experiments at various baselines. The dashed line corresponds to the classic 3 neutrino oscillation scenario, while the solid line corresponds to the 3+1 model (3 active neutrinos + 1 sterile neutrino).}
\label{fig:oer}
\end{figure}

An intriguing, nearly 3$\sigma$ indication of neutrino disappearance at baselines of less than 100 m has been recently revealed~\cite{bib:Mention}, as can be seen in Figure~\ref{fig:oer}. The indication appeared as a result of reanalysis of the global reactor antineutrino experimental data using the updated calculation of the reactor neutrino flux. A possible physical explanation for this effect, commonly called the Reactor Antineutrino Anomaly (RAA), may be the existence of yet-undiscovered mixing between the electron antineutrino and a new, 4$^{th}$ neutrino flavor that is ``sterile'', i.e., does not interact through the Standard Model weak interactions. To be compatible with the observed deficit, this sterile neutrino would need to have a mass splitting $|\Delta m^2_{new}|$ $\ge$ 0.1 eV$^2$. The RAA, in combination with further complementary hints from gallium solar neutrino experiments~\cite{bib:Gallex}, as well as the LSND and MiniBooNE experiments~\cite{bib:MiniBooNE}, create a compelling case to search for a sterile (4th) neutrino flavor and measure its mass. If detected, the presence of a 4$^{th}$-generation sterile neutrino would impact the extraction of the neutrino mixing angle $\theta_{13}$ from the reactor neutrino data, as well as future measurements of the CP-violating phase in the leptonic sector.
\section{Testing the Reactor Antineutrino Anomaly}
\begin{figure}[htb]
\centering
\includegraphics[scale=0.35]{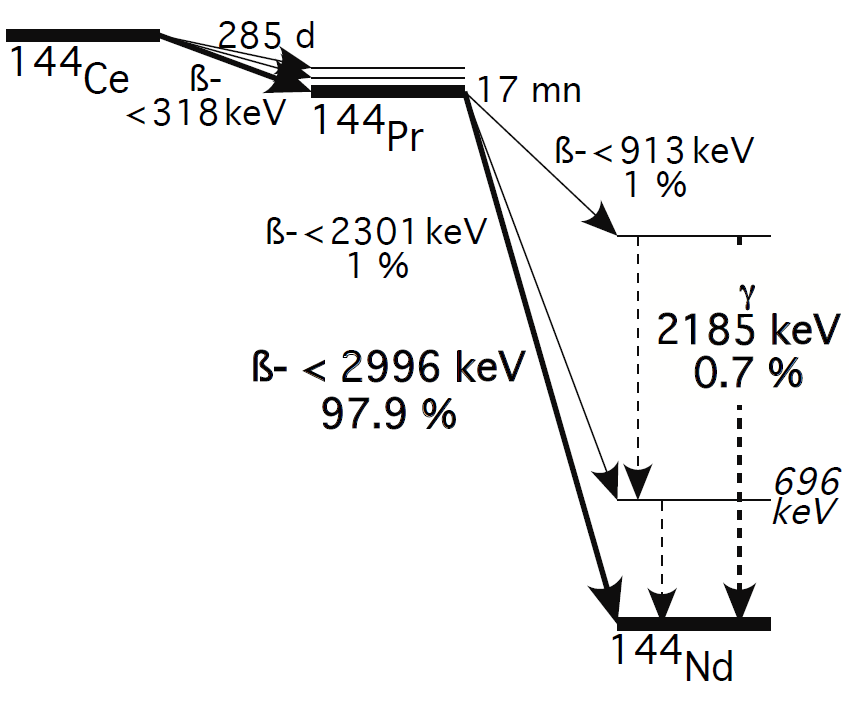}
\caption{$^{144}$Ce beta decays slowly with emission of low energy antineutrino below the IBD threshold. This decay is almost immediately followed by the beta decay of $^{144}$Pr with emission of antineutrino of energy up to 2.996 MeV. Antineutrinos in the high energy end of $^{144}$Pr decay spectrum, with energy above 1.8 MeV, can engage in IBD in LS. The dominant background is 2.185 MeV gamma produced with a branching fraction of 0.7\%.}
\label{fig:ds}
\end{figure}

A decisive test for the existence of the fourth neutrino flavor, as indicated by the RAA and gallium anomaly, can be achieved with an intense radioactive neutrino or antineutrino source deployed near a massive liquid scintillator detector. For neutrinos, monochromatic neutrino sources such as $^{51}$Cr or $^{37}$Ar can be used: the neutrino detection is realized through neutrino-electron scattering ($\nu + e^-$). Such measurements are quite challenging. In addition to the small $\nu+e^-$ cross-section, there are numerous backgrounds to this method including solar neutrinos and the decays of natural or cosmogenically induced radioactivity. To overcome these challenges, radioactive sources of 10 MCi or more are needed. The antineutrino sources benefit from the inverse beta decay (IBD) reaction with the characteristic prompt $e^+$ annihilation, followed by delayed neutron capture: such delayed-coincidence signature is an effective background suppression tool. We intend to use a 75 kCi $^{144}$Ce - $^{144}$Pr antineutrino source that produces two consecutive beta decays. The $^{144}$Ce - $^{144}$Pr antineutrino generator was first proposed in 2011~\cite{bib:cribier}. $^{144}$Ce, which has a half-life of 285 days, $\beta^-$ decays into a $^{144}$Pr daughter, which subsequently $\beta^-$ decays to the stable $^{144}$Nd with a Q-value of 2.996 MeV. A simplified decay scheme is shown in Figure~\ref{fig:ds}. The $^{144}$Pr decay produces antineutrinos with energy above the inverse beta decay threshold of 1.8 MeV. Given the short half-life of $^{144}$Pr (17.3 minutes), the $^{144}$Pr is always in equilibrium with $^{144}$Ce and the antineutrino flux depends solely on the decay rate of $^{144}$Ce.  The $^{144}$Ce source, in the form of CeO$_2$, will be produced at the Mayak reprocessing plant in Russia from 2-3 year old spent nuclear fuel. The $^{144}$Ce fraction in CeO$_2$ is ~0.5\% three years after the last irradiation of the fuel inside the reactor core.

The fourth neutrino flavor will reveal itself in the oscillatory pattern of the distance-dependent flux from a pure antineutrino source. The oscillation length should be approximately matched by the source-detector distance. Based on the implications of RAA, the best fit for the $|\Delta m^2_{new}| \sim 1$ eV$^2$ corresponds to the oscillatory length of roughly 1 meter. 

There are three candidate detectors in the world that are well suited for this source-based search for sterile neutrinos:  KamLAND, SNO+, and Borexino. This project, dubbed CeLAND, proposes to deploy a 75 kCi $^{144}$Ce source into KamLAND~\cite{bib:KL}. KamLAND is a 1 kton liquid scintillator detector that has been in operation since 2002. The target liquid scintillator is contained in the 13 m diameter balloon. The compactness of the source ($\sim$20 g and 13 cm in diameter and height), combined with possible 3-16 m long baselines, as well as the good energy and vertex position resolution of KamLAND, will provide a distance- and energy-dependent oscillation signature. In the proposed configuration, CeLAND will provide a definitive test of the existence of the fourth neutrino.

\begin{figure}[htb]
\centering
\includegraphics[scale=0.6]{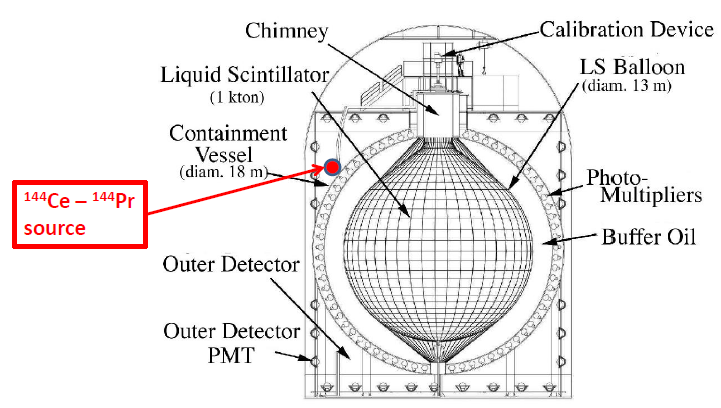}
\caption{Schematic representation of the KamLAND detector and approximate location of the cerium source inside the OD. The location is easily accessible through a wide hatch on the top of the OD.}
\label{fig:cl}
\end{figure}

\begin{figure}[htb]
\centering
\includegraphics[scale=0.4]{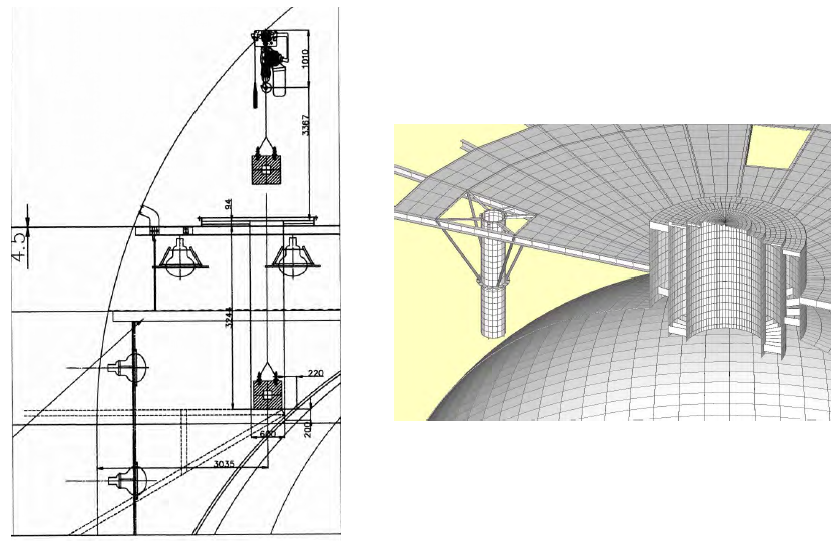}
\caption{The preliminary sock design to contain the cerium source with shielding inside KamLAND OD. The source will be located at the bottom of the cylindrical sock, at approximate distance of 9.5 m from the detector center.}
\label{fig:sock}
\end{figure}

The $^{144}$Ce source will be placed in the KamLAND Outer Detector (OD) region, close to the outside surface of the stainless steel sphere that encloses the KamLAND Inner Detector (ID), at the approximate distance of 9.5 m from the detector center. The source with the shielding will be placed in a cylindrical structure vessel called the sock, supported from the top of the OD.
Figure~\ref{fig:cl} shows approximate location of the source inside the KamLAND OD detector. Figure~\ref{fig:sock} shows preliminary sock design that will hold the source. The deployment in the OD will probe baseline range from 3 m (closest part of the target LS filled balloon) up to 16 m (the opposite side of the target LS filled balloon), expanding the sterile neutrino mixing parameter space that can be probed.  Deployment in the OD will enable CeLAND to run in parallel with KamLAND-Zen (the $^{136}$Xe $\beta \beta 0 \nu$ search~\cite{bib:KLzen}). In this configuration, CeLAND will be able to probe the RAA parameter space with $>$ 95\% C.L. The $^{144}$Ce source run in the OD will last 1.5 years.

Design and construction of adequate tungsten shielding for a 75 kCi radioactive source will be essential to background reduction as well as for safe transport from Mayak in Russia to Kamioka mine in Japan. Various transport scenarios are under investigation. The shielding serves a dual purpose: reduce the radiation from the $^{144}$Ce source to the safe level for transportation, handling, and human presence; and suppress beta and gamma backgrounds in the target LS during the source run in KamLAND. The deployment in the OD takes additional advantage of shielding from 11.5 mm thick wall of stainless steel that encloses the ID.              The 2.5 m thick layer of mineral oil inside the ID that surrounds the balloon containing target LS, provides further shielding from the $^{144}$Ce source beta and gamma emissions. In addition, shielding made of a dense tungsten alloy will be designed for biological protection and suppression of the $^{144}$Pr 2.185 MeV gamma radiation that is a dangerous background for neutron capture on hydrogen. Preliminary estimates show that 16 cm thick tungsten shield will be sufficient to provide both biological protection and background suppression.

Precise determination of the absolute activity of a 75 kCi radioactive source at 1-1.5\% level as in~\cite{bib:Gallex} will further enhance the sensitivity of the measurement. A calorimeter will be built in order to measure the absolute activity of the source.

\section{Sensitivity and prospects}
\begin{figure}[htb]
\centering
\begin{minipage}{.5\textwidth}
  \centering
  \includegraphics[width=0.95\linewidth]{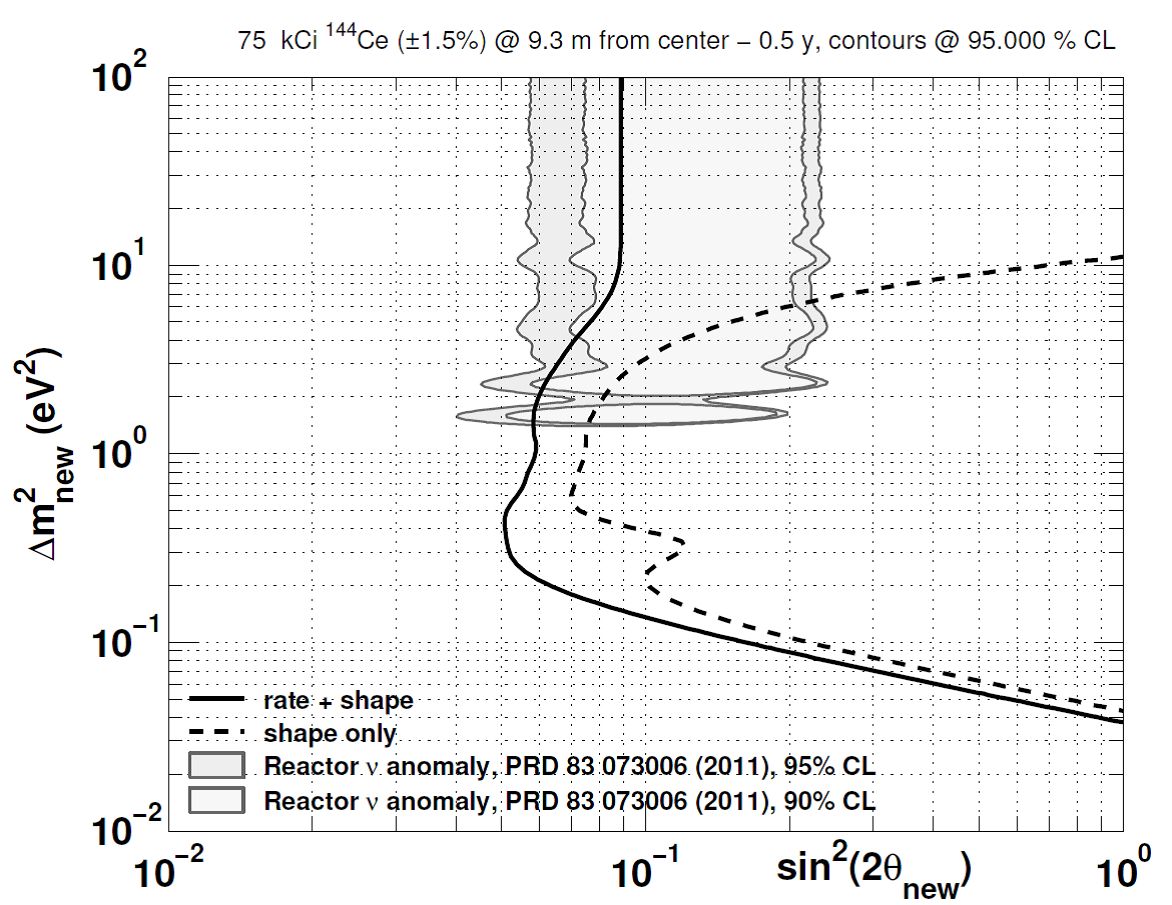}
\end{minipage}%
\begin{minipage}{.5\textwidth}
  \centering
  \includegraphics[width=0.95\linewidth]{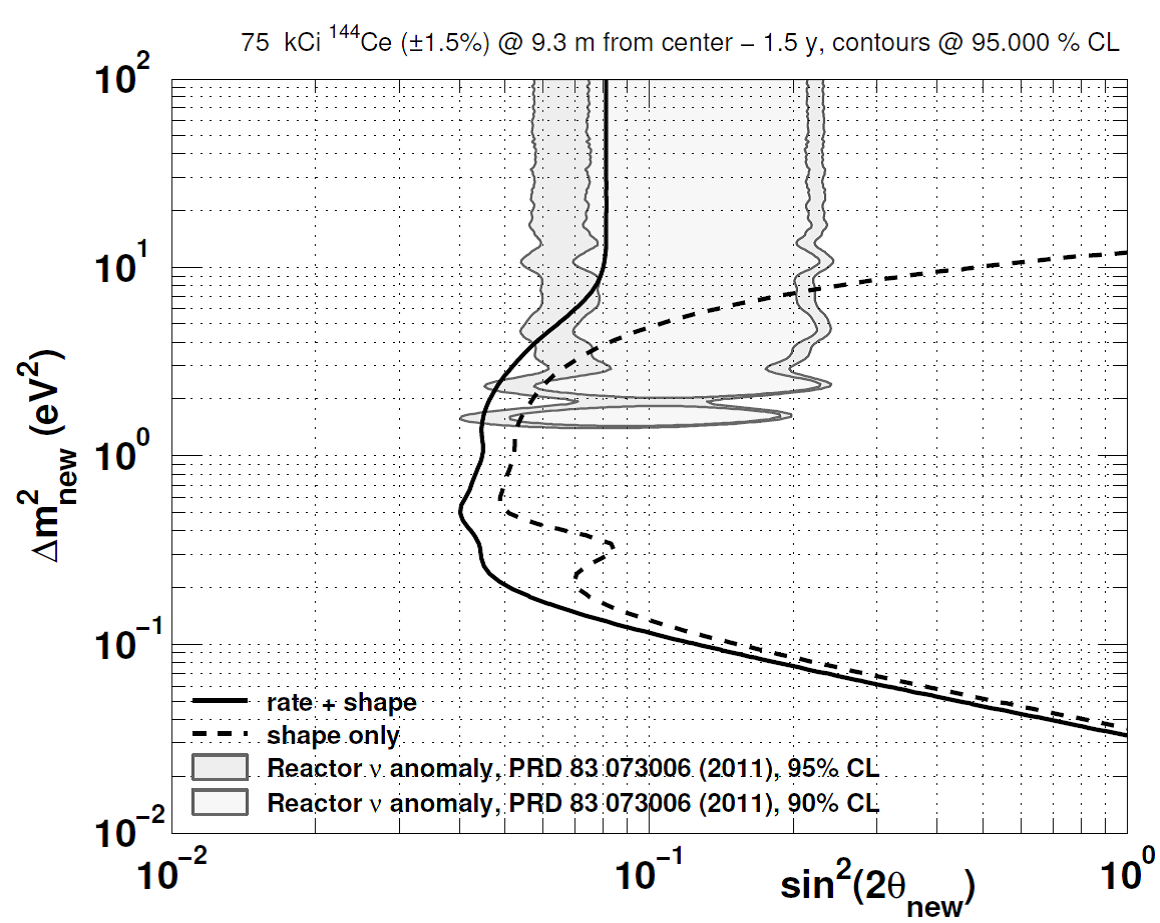}
\end{minipage}
\caption{Sensitivity of the measurement to exclude the non-oscillation after (left) 6 months and (right) 18 months of data taking in the KamLAND OD is shown. 75 kCi $^{144}$Ce source, located 9.5 m from the detector center is assumed. Statistical sample with fiducial volume cut at 6 m radius is used. The light and dark gray area show 90\% C.L. and 95\% C.L. limits of RAA. The dashed line corresponds to the 95\% C.L. non-oscillation exclusion with energy spectrum shape analysis only. The solid line corresponds to the 95\% C.L. non-oscillation hypothesis exclusion with both absolute event rate and event energy spectrum shape analysis. The inclusion of the rate information significantly improves the sensitivity, in particular in the large $\Delta m^2_{new}$ region.}
\label{fig:sensitivity}
\end{figure}

In one and half year of data taking CeLAND will probe the majority of the phase space covered by the RAA at $>$ 95\% C.L. Figure~\ref{fig:sensitivity} shows the sensitivity to short baseline oscillations after 6 and 18 months of data taking in the OD. While the shape only analysis can probe the region where the current best fit solution is located, inclusion of the rate analysis, (assuming 1.5\% absolute source activity uncertainty) significantly increases the coverage of the region of interest, indicated by the RAA. Even with just 6 months of data, significant limit in the area of interest can be placed. 

At the end of the 144Ce source run in the OD after 18 months, the sensitivity increases significantly as can be seen in the Figure~\ref{fig:sensitivity}. 95\% C.L. non-oscillation hypothesis exclusion is achieved for the mixing angles as low as $\sin^2 2\theta_{new}$ $\sim$ 0.05 in the region of interest and for high values of $\Delta m^2_{new}$. Excellent sensitivity to the parameter space with small $\Delta m^2_{new} \sim$ 0.1 eV$^2$ is achieved due to relatively long baseline applied. It is evident from the plots that the accurate measurement of the absolute activity of the $^{144}$Ce source is essential.

\begin{figure}[h!tb]
\centering
\includegraphics[scale=0.5]{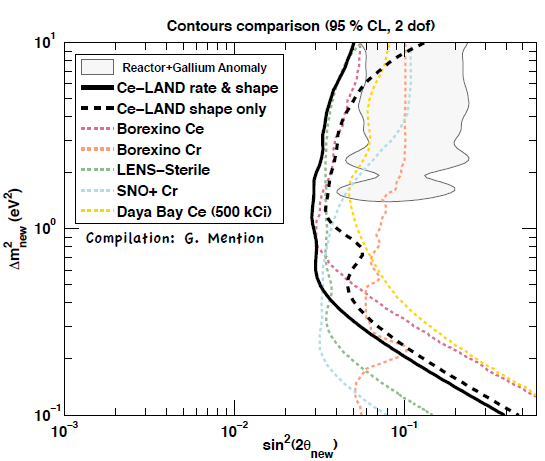}
\caption{Sensitivity limits of several proposed source experiments. CeLAND’s combined shape and rate sensitivity exceeds other similar proposed experiments~\cite{bib:lasserre}.}
\label{fig:comparison}
\end{figure}

Thanks to the energy and vertex information of the $^{144}$Ce - $^{144}$Pr antineutrino spectrum, and KamLAND’s large size, CeLAND has an excellent sensitivity compared to other similar proposed experiments~\cite{bib:lasserre}, as can be seen in Figure~\ref{fig:comparison}. 

The CeLAND experiment provides an effective, fast, and inexpensive way to search for sterile neutrinos at the short baseline, covering the RAA phase space completely at the 95\% C.L. It is complementary to proposed searches with reactor- and accelerator-produced neutrinos. If CeLAND demonstrates the existence of sterile neutrinos, it would be one of the most significant discoveries of physics beyond the Standard Model. The measurement could also potentially influence the precision determination of $\theta_{13}$, and consequently impact prospects for the measurement of CP-violation in the lepton sector. A null measurement will definitively put to rest sterile neutrino interpretations of the existing neutrino anomalies.

\end{document}